\documentclass[a4paper]{spie}  
\usepackage{amsmath}
\usepackage{amsfonts}
\usepackage[dvips]{graphicx}

\title{Photon counting schemes and performance of non--deterministic
  nonlinear gates in linear optics} 

\author{S. D. Bartlett\supit{a}, E. Diamanti\supit{b}, B.
  C. Sanders\supit{a,b}, and Y. Yamamoto\supit{b}
\skiplinehalf
\supit{a}Department of Physics and
        Centre for Advanced Computing -- Algorithms and Cryptography,   \\
        Macquarie University, Sydney, New South Wales 2109, Australia\\
\supit{b}Quantum Entanglement Project, ICORP, JST,
        Edward L.\ Ginzton Laboratory, \\ Stanford University, 
        California 94305-4085
}

\authorinfo{Further author information: (Send correspondence to
  B.C.S.)\\B.C.S.: E-mail: barry.sanders@mq.edu.au, Telephone: +61 2
  9850 8935}

\begin{document} 
  \maketitle 

\begin{abstract}
  The performance of nondeterministic nonlinear gates in linear
  optics relies on the photon counting scheme being employed and the
  efficiencies of the detectors in such schemes. We assess the
  performance of the nonlinear sign gate, which is a critical
  component of linear optical quantum computing, for two standard
  photon counting methods: the double detector array and the visible
  light photon counter. Our analysis shows that the double detector
  array is insufficient to provide the photon counting capability for
  effective nondeterministic nonlinear transformations, and we
  determine the gate fidelity for both photon counting methods as a
  function of detector efficiencies.
\end{abstract}



\section{Introduction}

Nonlinear transformations in quantum optics can yield dramatic
nonclassical effects, particularly if the Kerr optical nonlinearity
(intensity--dependent refractive index) is
exploited~\cite{Mil86,Yur85,Buz95}.  In the context of optical quantum
information theory, this nonlinearity has been identified as an
important resource for processing photons encoded via the dual rail
approach~\cite{Chu95}; in practice, though, nonlinear materials suffer
from weak strengths and high losses.  Recently Knill, Laflamme and
Milburn (KLM)~\cite{KLM01} showed that a nondeterministic Kerr
transformation can be effected for states with up to two photons via
linear optical interferometry, and the cases where the desired
transformation has been performed are distinguished by photon counting
measurements on ancilla modes.  Although this nondeterministic
nonlinear transformation works only a fraction of the time, the gate
operation can be used offline from the quantum computation, and the
cases where the gate has operated properly can be incorporated into
the quantum circuit via quantum teleportation~\cite{Got99}.  The
roadmap to basic linear optics quantum computation~\cite{KLM00} begins
with the implementation of a nondeterministic nonlinear sign change.
This gate is also of great interest for quantum optics applications
that require nonlinear transformations of states that can be written
as a superposition of a finite number of Fock states;
e.g.~\cite{Kok01a}

Linear optics transformations involve beam splitters, phase shifters
and mirrors, and these devices are well understood~\cite{Vog94}, but
the other key aspect of the nondeterministic nonlinear gate is the
employment of photon counters.  This requirement of photon counting is
distinct from photodetection, which generally detects only the
presence or absence of light~\cite{Res01,Bar02}; photon counting
measurements must be able to detect the number of indistinguishable
photons in the same spatial, temporal, and polarization mode.  With
this in mind, KLM discuss practical means for counting photons via an
array of beamsplitters and single--photon counting modules (SPCMs),
incorporating the effects of finite--efficiency photodetection.  The
use of arrays of SPCMs to construct a photon counter has been studied
in detail by Kok and Braunstein~\cite{Kok01b}.

Photodetection plays a critical role in two stages of the KLM scheme.
The first stage is the nondeterministic gate that performs the
nonlinear transformation (specifically, the nonlinear sign (NS) gate),
and the success or failure is determined by photon counting of
ancillary modes.  The second stage is the quantum teleportation of a
successful gate operation, where photon counting also plays an
important role.  Our interest here is the nonlinear transformation;
the effects of limited photon detection efficiency in the latter stage
of quantum teleportation has been investigated by Glancy \emph{et
  al}~\cite{Gla02} who show that a high efficiency is required for
teleportation of the gate to succeed.  In this paper, we consider the
NS gate operation alone, and we use the gate fidelity~\cite{Nie00} to
characterise successful nonlinear transformations.  Ultimately the
success or failure of the transformation depends on the application of
this component of a quantum optical circuit, but the fidelity provides
a useful measure of the gate's performance.  As a reference, we
present the gate fidelity for a deterministic linear optical
transformation and point out that, if the gate fidelity of the NS gate
is less than that of the deterministic linear one, it is essentially
useless at performing a nonlinear transformation.  Consequently the
gate fidelity for the deterministic linear optical transformation
gives a lower bound for a gate's capability of performing a nonlinear
transformation.

In practice, a very large array of SPCMs is not feasible to perform
the photon counting measurement.  In Section~\ref{sec:measurements} we
describe two measurement schemes, the double detector array (DDA) and
the visible light photon counter (VLPC)~\cite{Kim99,Tak99}, which
comprise realistic schemes to approximate a photon counting
measurement.  We analyse the role of the DDA and the VLPC in
nondeterministic nonlinear transformations in detail in
Section~\ref{sec:NSGate}, focussing on the NS gate of the KLM scheme
with particular attention directed towards a proper accounting of
photodetection efficiency.  We calculate the gate fidelity for each of
two cases (for the DDA and for the VLPC) as a function of detector
efficiency and compare to that of a linear optical transformation.  We
conclude, in Section~\ref{sec:Conclusions}, with how current detectors
and technology may be used to perform nondeterministic nonlinear
transformations in quantum information and quantum optics.

\section{Photon counting measurements}
\label{sec:measurements}

Photon counting is the measurement of the number of photons in a mode
(i.e., the number of intrinsically indistinguishable photons
propagating in the same temporal, spatial, and polarization mode).
Such a measurement is described by the \emph{photon counting
  POVM}~\cite{Bar02}:
\begin{equation}
  \label{eq:PhotonCountingPOVM}
  \bigl\{ \Pi_n = |n\rangle\langle n|, n=0,1,\ldots \bigr\} \, ,
\end{equation}
where $|n\rangle$ is the $n$--photon Fock state.

Such a measurement cannot be achieved using current
technology~\cite{Bar02}.  In the following, we describe two
measurement schemes that can, to some degree, approximate this
measurement: the double detector array (DDA) and the visible light
photon counter (VLPC).

\subsection{Detector cascades}

In general, photodetection operates on a threshold principle: an event
is registered if any light is detected, and the number of quanta is
not revealed in this signal.  We define a \emph{threshold detector} as
a detector that is triggered by one or more photons, and cannot
distinguish the number of photons~\cite{Bar02}.  Thus, the POVM for a
threshold detector consists of two elements, $\Pi_0$ and $\Pi_{>0}$,
corresponding to the detection of no photons or at least one photon.
For a threshold detector with unit efficiency, the POVM is
\begin{equation}
  \label{eq:UnitEfficiencyTD}
  \Pi_0^{\eta=1} = |0\rangle \langle 0| \, , \qquad
  \Pi_{>0}^{\eta=1} = \sum_{n=1}^{\infty} |n\rangle \langle n| \, .
\end{equation}
With finite detection efficiency $\eta<1$ (where $\eta$ is the
probability of registering a detection given an input of a one--photon
Fock state), the POVM is
\begin{equation}
  \label{eq:FiniteEfficiencyTD}
  \Pi_0^{\eta} = \sum_{n=0}^{\infty} (1-\eta)^n |n\rangle \langle n|
  \, , \qquad 
  \Pi_{>0}^{\eta} = \sum_{n=0}^{\infty} \bigl[1-(1-\eta)^n \bigr] |n\rangle
  \langle n| \, .
\end{equation}

\begin{figure}
  \includegraphics*[width=6in,keepaspectratio]{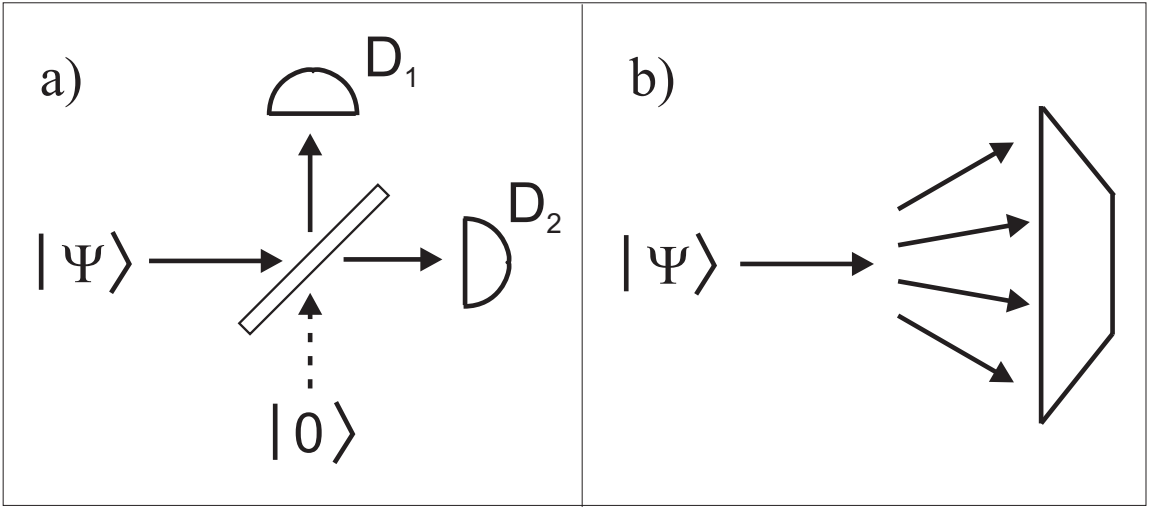}
  \caption{Diagrammatic representation of the two photon counting
    models considered in this paper.  a) The DDA, involving two
    detectors $D_1$ and $D_2$ on the output modes of a beamsplitter.
    b) The VLPC, which acts by dispersing the input mode over several
    independent detector modes.}
  \label{fig:DDAVLPC}
\end{figure}

Using a 50/50 beamsplitter and two finite efficiency threshold
detectors, one can construct a detector cascade~\cite{Kok01b} that can
(sometimes) distinguish one photon from two; we refer to such a scheme
as a double detector array (DDA).  The state to be measured is
incident on a 50/50 beamsplitter with the vacuum state injected at the
other input, and a finite efficiency threshold detector is placed at
each output; see Fig.~\ref{fig:DDAVLPC}a.
The POVM for the DDA consists of three elements, $\Pi_{\text{none}}$,
$\Pi_{\text{one}}$ and $\Pi_{\text{both}}$, corresponding to whether
none, one, or both of the threshold detectors has detected at least
one photon.  (Note that we do not distinguish which of the two
detectors has fired, as it does not provide any information about the
input state.)  This POVM is given by
\begin{align}
  \label{eq:2TDCascadePOVM}
  \Pi_{\text{none}} &= \sum_{n=0}^{\infty} (1-\eta)^n |n\rangle
  \langle n| \, , \nonumber \\
  \Pi_{\text{one}} &= \sum_{n=0}^\infty 2 \bigl[ (1-\eta/2)^n -
  (1-\eta)^n \bigr] |n\rangle \langle n| \, , \nonumber \\
  \Pi_{\text{both}} &= \sum_{n=0}^\infty \bigl[ 1 + (1-\eta)^n -
  2(1-\eta/2)^n \bigr] |n\rangle \langle n| \, .  
\end{align}
Detector cascades can be designed to distribute the photons equally
over $N>2$ modes using a multisplitter array and $N$ threshold
detectors~\cite{Kok01b}, but such a scheme becomes progressively more
difficult to implement as $N$ increases.

\subsection{Visible light photon counter}

The VLPC~\cite{Kim99,Tak99} functions differently than a threshold
detector.  The incoming mode is dispersed over many detector modes
(active regions) that behave independently; see
Fig.~\ref{fig:DDAVLPC}b.  The VLPC can be modelled as a detector
cascade using $N$ beamsplitters and finite efficiency threshold
detectors.  For the low photon numbers used in the KLM scheme, the
probability of two or more photons entering the same active region is
negligible; thus, one can use the approximation $N \to \infty$.  The
POVM for the VLPC consists of an infinite number of elements $\{
\Pi_k, k=0,1,2,\ldots \}$ corresponding to detecting $k$ photons.
This POVM is given by
\begin{equation}
  \label{eq:VLPCPOVM}
  \Pi_k = \sum_{n=k}^\infty \binom{n}{k} \eta^k (1-\eta)^{n-k}
  |n\rangle \langle n| \, .
\end{equation}
For small $k$, this POVM agrees with experimental tests of the
VLPC~\cite{Kim99,Edo02}.

\section{Nonlinear sign gate}
\label{sec:NSGate}

Consider a quantum state of a radiation mode to be of the form
\begin{equation}
  \label{eq:StateUpTo2}
  |\psi\rangle = \alpha |0\rangle + \beta |1\rangle + \gamma|2\rangle
   \, ,
\end{equation}
with $|\alpha|^2 + |\beta|^2 + |\gamma|^2 = 1$.  By interacting this
mode with a Kerr medium, described by an interaction
Hamiltonian~\cite{Wal94}
\begin{equation}
  \label{eq:KerrHam}
  \hat{H}_{\rm Kerr} = \frac{\chi}{2} (\hat{a}^\dag)^2 (\hat{a})^2 \, ,
\end{equation}
the state is transformed as
\begin{equation}
  \label{eq:DeterministicKerrTrans}
  \exp(i\hat{H}_{\rm Kerr}t): |\psi\rangle \to \alpha |0\rangle +
  \beta |1\rangle + \exp(i\chi t) \gamma |2\rangle \, .
\end{equation}
For interaction time $t = \pi/\chi$, the resulting transformation is a
sign change ($\pi$ phase shift) on the two--photon component.

\begin{figure}
  \includegraphics*[width=6in,keepaspectratio]{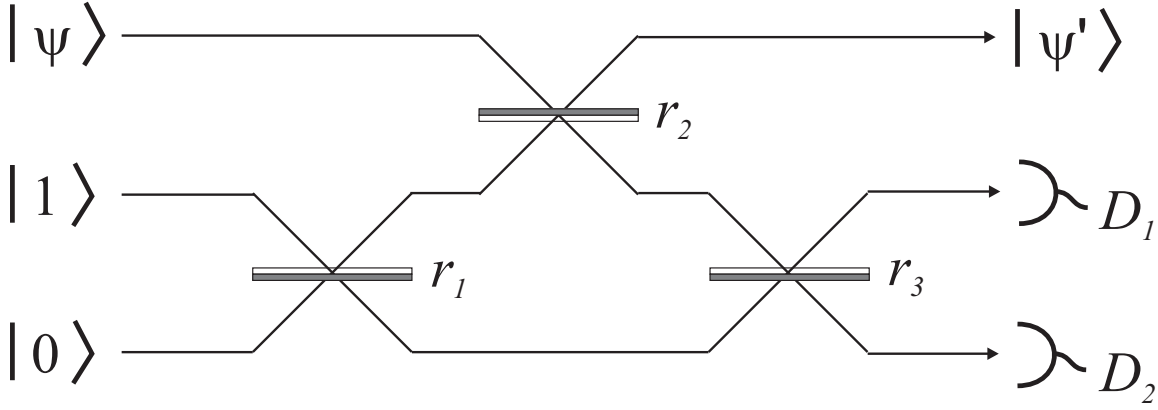}
  \caption{Diagrammatic representation of the NS gate.  The intensity
    reflectivities of the beamsplitters are $r_1 = r_3 =
    (4-2\sqrt{2})^{-1}$ and $r_2 = (\sqrt{2}-1)^2$, and the shaded
    surface indicates the surface where a sign change takes place.
    The gate reports success when detector $D_1$ measures exactly one
    photon, and detector $D_2$ measures zero.}
  \label{fig:NSGate}
\end{figure}

Using only linear optics and photon counting, the nonlinear sign (NS)
gate of KLM can implement this nonlinear transformation, albeit in a
nondeterministic fashion.  The design of the gate is presented in
Fig.~\ref{fig:NSGate}; see~\cite{Ral02} for details.  
Two auxilliary modes are used; a single photon is injected into one, and
the vacuum state into the other.  If a single photon is detected at
detector~$1$ and no photon is detected at detector~$2$, the resulting
transformation on the (unknown) input state is given by
\begin{equation}
  \label{eq:NSideal}
  {\rm NS}:  |\psi\rangle = \alpha|0\rangle + \beta|1\rangle +
   \gamma|2\rangle  
   \to |\psi'\rangle = \frac{1}{2} \bigl( \alpha|0\rangle + \beta|1\rangle -
   \gamma|2\rangle \bigr) \, .
\end{equation}
The normalization factor of $1/2$ represents the fact that
this transformation is achieved only one quarter of the time (the
resulting state has square modulus $1/4$).

The NS gate requires photon counters that can discriminate the number
of photons; the measurements must yield \emph{exactly} one photon at
detector~$1$ and none at detector~$2$.  With ideal photon counters
described by the photon counting POVM of
Eq.~(\ref{eq:PhotonCountingPOVM}), the NS gate implements the
transformation of Eq.~(\ref{eq:NSideal}).  However, with either of the
realistic photon counting schemes discussed in the previous section,
the resulting transformation will be different.  Photodetectors will
miss events that should be kept, and they will also miscount photons
and allow events that should be discarded.  Thus, the output of the NS
gate will be a mixed state which depends on the parameters of the
photon counting measurements.  In the following, we calculate the
resulting mixed state output of an NS gate employing either of the
photon counting schemes of the previous section.

\subsection{NS gate using photon counting models}

Notation is defined as follows.  Let the input state $|\psi\rangle$
enter mode~$1$, the single photon state $|1\rangle$ enter mode~$2$
and the vacuum $|0\rangle$ enter mode~$3$.  The three--mode state is
passed through a sequence of three beamsplitters with intensity
reflectivity $r_1$, $r_2$ and $r_3$ respectively.  The transformation
given by these beamsplitters can be expressed in the form
\begin{multline}
  \label{eq:BeamsplitterTrans}
  \bigl( \alpha |0\rangle_1 + \beta|1\rangle_1 + \gamma |2\rangle_1
  \bigr) |1\rangle_2 |0\rangle_3 \\
  \to \alpha \sum_{i+j+k=1} c_{ijk} |i\rangle_1 |j\rangle_2
  |k\rangle_3 
  + \beta \sum_{i+j+k=2} c_{ijk} |i\rangle_1 |j\rangle_2
  |k\rangle_3 
  + \gamma \sum_{i+j+k=3} c_{ijk} |i\rangle_1 |j\rangle_2
  |k\rangle_3  \, ,
\end{multline}
where $c_{ijk}$ be the complex probability amplitude of $i,j,k$
photons exiting modes $1,2,3$, respectively.  Photon number
measurements are performed on modes~$1$ and $2$, and the output state
exits mode~$3$ conditioned on a measurement of zero photons in mode~$1$
and one photon in mode~$2$.  The reflectivities are chosen to be
\begin{equation}
  \label{eq:Reflectivities}
  r_1 = r_3 = (4-2\sqrt{2})^{-1} \, , \quad
  r_2 = (\sqrt{2}-1)^2 \, ,
\end{equation}
which gives $c_{010} = c_{011} = 1/2$ and $c_{012} = -1/2$.  Note that
this condition gives $c_{110} = 0$ (an additional simplification).

The other probability amplitudes for contributing Fock state outputs
as functions of intensity reflectivities of the beamsplitters $r_1$,
$r_2$ and $r_3$ are
\begin{align}
  \label{eq:ProbAmplitudes}
  c_{111} &= -\sqrt{2(1-r_1)r_2(1-r_2)}(1-2r_3) +
  \sqrt{2r_1(1-r_2)r_3(1-r_3)}(1-3r_2) \nonumber \\ 
  c_{020} &= \sqrt{2r_1r_2(1-r_2)}r_3 +
  \sqrt{2(1-r_1)(1-r_2)r_3(1-r_3)} \nonumber \\  
  c_{021} &= -2\sqrt{(1-r_1)r_2(1-r_2)r_3(1-r_3)} +
  \sqrt{r_1(1-r_2)}(1-3r_2)r_3 \nonumber \\ 
  c_{210} &= 3\sqrt{r_1r_2r_3}(1-r_2)(1-r_3) +
  \sqrt{(1-r_1)(1-r_3)}(1-r_2)(1-3r_3) \nonumber \\ 
  c_{120} &= 3\sqrt{r_1r_2(1-r_3)}(1-r_2)r_3 +
  \sqrt{(1-r_1)r_3}(1-r_2)(2-3r_3) \nonumber \\  
  c_{030} &= \sqrt{3r_1r_2}(1-r_2)(\sqrt{r_3})^3 +
  \sqrt{3(1-r_1)(1-r_3)}(1-r_2)r_3 \, . 
\end{align}

To calculate the (generally mixed) output state of a NS gate with a
given detection scheme, one projects the three--mode state with the
POVM elements for the detection of zero photons in mode $1$ and one
photon in mode $2$, then normalizes by the probability of this
measurement.  First, we consider a NS gate employing DDAs to perform
the photon number measurements.  A DDA, modelled by the POVM of
Eq.~(\ref{eq:2TDCascadePOVM}), is placed at each of the output ports
of modes $1$ and $2$.  The resulting unnormalized density matrix
conditioned on a measurement of zero photons in mode $1$ and one
photon in mode $2$ is given by
\begin{multline}
  \label{eq:rhoN=2TDC}
  \bar{\rho}'_{\rm DDA} = \eta|\psi'\rangle\langle\psi'| 
  + \eta(1-\eta)|\phi_1\rangle\langle\phi_1| + [\tfrac{1}{2}\eta^2 +
  2\eta(1-\eta)] |\phi_2\rangle\langle\phi_2| 
  + \eta(1-\eta)^2 |\phi_3\rangle\langle\phi_3| \\
  + (1-\eta)[\tfrac{1}{2}\eta^2 +
  2\eta(1-\eta)] |\phi_4\rangle\langle\phi_4| 
  + 2[(1-\eta/2)^3 - (1-\eta)^3]|\phi_5\rangle\langle\phi_5| \, ,
\end{multline}
with unnormalized states
\begin{equation}
  \label{eq:N=2TDCextraterms}
  |\phi_1\rangle = \gamma c_{111} |1\rangle
   \, , \quad
  |\phi_2\rangle = \beta c_{020}|0\rangle + \gamma c_{021} |1\rangle
   \, , \quad
  |\phi_3\rangle = \gamma c_{210} |0\rangle \, , \quad
  |\phi_4\rangle = \gamma c_{120} |0\rangle \, , \quad
  |\phi_5\rangle = \gamma c_{030} |0\rangle \,.
\end{equation}
The trace of this density matrix is
\begin{multline}
  \label{eq:N=2TDCTrace}
  {\rm Tr}(\bar{\rho}'_{\rm DDA}) = \tfrac{1}{4} \eta + \eta(1-\eta)|\gamma
  c_{111}|^2 
  +[\tfrac{1}{2}\eta^2 + 2\eta(1-\eta)](|\beta c_{020}|^2 + |\gamma
  c_{021}|^2) 
  + \eta(1-\eta)^2 |\gamma c_{210}|^2 \\
  + (1-\eta)[\tfrac{1}{2}\eta^2 + 2\eta(1-\eta)]|\gamma c_{120}|^2 
  + 2[(1-\eta/2)^3 - (1-\eta)^3] |\gamma c_{030}|^2 \, .
\end{multline}
This trace gives the probability of an apparent success; i.e., the
probability of a measurement of zero photons at mode $1$ and one
photon at mode $2$.  Note that ``apparent success'' is used; by that,
we mean that the photon counters report zero and one photons
respectively, although this does \emph{not} imply that the gate has
succesfully implemented the transformation (\ref{eq:NSideal}).  The
normalized density matrix for this scheme is then given by
\begin{equation}
  \label{eq:DDArhonorm}
  \rho'_{\rm DDA} = \frac{\bar{\rho}'_{\rm DDA}}{{\rm Tr}(\bar{\rho}'_{\rm
  DDA})} \, .
\end{equation}

Using the VLPC to perform the photon measurements gives different
results.  The resulting unnormalized density matrix conditioned on a
measurement of zero photons in mode $1$ and one photon in mode $2$ is
given by
\begin{multline}
  \label{eq:rhoVLPC}
  \bar{\rho}'_{\rm VLPC} = \eta|\psi'\rangle\langle\psi'| 
  + \eta(1-\eta)|\phi_1\rangle\langle\phi_1| +  2\eta(1-\eta)
  |\phi_2\rangle\langle\phi_2|  \\
  + \eta(1-\eta)^2 \bigl( |\phi_3\rangle\langle\phi_3|  
  + 2 |\phi_4\rangle\langle\phi_4| \bigr) 
  + 3\eta(1-\eta)^2 |\phi_5\rangle\langle\phi_5| \, .
\end{multline}
The trace of this density matrix gives the apparent success
probability:
\begin{multline}
  \label{eq:VLPCTrace}
  {\rm Tr}(\bar{\rho}'_{\rm VLPC}) = \tfrac{1}{4} \eta + \eta(1-\eta)|\gamma
  c_{111}|^2 
  + 2\eta(1-\eta)(|\beta c_{020}|^2 + |\gamma c_{021}|^2) \\
  + \eta(1-\eta)^2 \bigl( |\gamma c_{210}|^2 + 2|\gamma c_{120}|^2\bigr) 
  + 3\eta(1-\eta)^2 |\gamma c_{030}|^2 \, .
\end{multline}
The resulting normalized density matrix for the VLPC--based scheme is
thus
\begin{equation}
  \label{eq:VLPCrhonorm}
  \rho'_{\rm VLPC} = \frac{\bar{\rho}'_{\rm VLPC}}{{\rm Tr}(\bar{\rho}'_{\rm
  VLPC})} \, .
\end{equation}

\subsection{Fidelity and Gate Fidelity}

With ideal photodetection, the output state of the NS gate is
$|\psi'\rangle = \alpha|0\rangle + \beta|1\rangle - \gamma|2\rangle$;
with the above photodetection models, the output is instead a mixed
state given by Eq.~(\ref{eq:rhoN=2TDC}) or (\ref{eq:rhoVLPC}).  We can
compare the desired outcome $|\psi'\rangle$ to these mixed states
directly by employing the fidelity, which characterises the
distinguishability of two states.  Calculating the fidelity between a
pure state $|\psi'\rangle$ and a mixed state $\rho'$ is
straightforward, and given by
\begin{equation}
  \label{eq:FidelityDef}
  F(|\psi'\rangle,\rho') = \sqrt{\langle \psi' | \rho' |\psi'\rangle} \, .
\end{equation}

We calculate the fidelity of the output states $\rho'_{\rm DDA}$ and
$\rho'_{\rm VLPC}$ for the DDA-- and VLPC--based schemes,
respectively.  For simplicity, we calculate the fidelity of the
unnormalized states, given by Eqs.~(\ref{eq:rhoN=2TDC}) and
(\ref{eq:rhoVLPC}), and then normalize the fidelity appropriately.
For the DDA--based scheme, we find 
\begin{multline}
  \label{eq:DDAOverlap}
  \langle \psi'| \bar{\rho}'_{\rm DDA} | \psi'\rangle = \tfrac{1}{4} \eta +
  \eta(1-\eta)|\beta^* \gamma c_{111}|^2 
  +[\tfrac{1}{2}\eta^2 + 2\eta(1-\eta)](|\alpha^* \beta c_{020} +
  \beta^* \gamma c_{021}|^2) \\
  + \eta(1-\eta)^2 |\alpha^* \gamma c_{210}|^2 
  + (1-\eta)[\tfrac{1}{2}\eta^2 + 2\eta(1-\eta)]|\alpha^* \gamma
  c_{120}|^2 
  + 2[(1-\eta/2)^3 - (1-\eta)^3] |\alpha^* \gamma c_{030}|^2 \, ,
\end{multline}
and thus,
\begin{equation}
  \label{eq:DDAFidelity}
  F(|\psi'\rangle,\rho'_{\rm DDA}) =
  \sqrt{\frac{\langle \psi'| \bar{\rho}'_{\rm DDA} |
  \psi'\rangle}{{\rm Tr}(\bar{\rho}'_{\rm DDA})}} \, . 
\end{equation}
For the VLPC--based scheme, we find
\begin{multline}
  \label{eq:VLPCOverlap}
  \langle \psi'| \bar{\rho}'_{\rm VLPC} | \psi'\rangle = \tfrac{1}{4}
  \eta +  \eta(1-\eta)|\beta^* \gamma c_{111}|^2 
  + 2\eta(1-\eta)(|\alpha^* \beta c_{020} +
  \beta^* \gamma c_{021}|^2) 
  + \eta(1-\eta)^2 |\alpha^* \gamma c_{210}|^2 \\
  + 2\eta(1-\eta)^2 |\alpha^* \gamma c_{120}|^2  
  + 3\eta(1-\eta)^2 |\alpha^* \gamma c_{030}|^2 \, ,
\end{multline}
and that the fidelity is given by
\begin{equation}
  \label{eq:VLPCFidelity}
  F(|\psi'\rangle,\rho'_{\rm VLPC}) =
  \sqrt{\frac{\langle \psi'| \bar{\rho}'_{\rm VLPC} |
  \psi'\rangle}{{\rm Tr}(\bar{\rho}'_{\rm VLPC})}} \, . 
\end{equation}

The gate fidelity~\cite{Nie00} (specifically, for the NS gate
discussed here) can be defined by the worst case scenario, i.e.,
\begin{equation}
  \label{eq:GateFidelityDef}
  F_{\rm NS gate} = \min_{|\psi\rangle} F (|\psi'\rangle,
  \rho') \,.
\end{equation}
The overlap for both detection schemes, as given by
Eqs.~(\ref{eq:DDAOverlap}) and (\ref{eq:VLPCOverlap}), is
minimised for any of the three Fock states $|0\rangle$, $|1\rangle$,
or $|2\rangle$.  The apparent success rate for both schemes, given by
Eqs.~(\ref{eq:N=2TDCTrace}) and (\ref{eq:VLPCTrace}), is maximised
for the state $|1\rangle$.  Thus, both photodetection models possess a
gate fidelity determined by the input state $|\psi\rangle =
|1\rangle$, which corresponds to $\beta = 1$.

\begin{figure}
  \includegraphics*[width=6in,keepaspectratio]{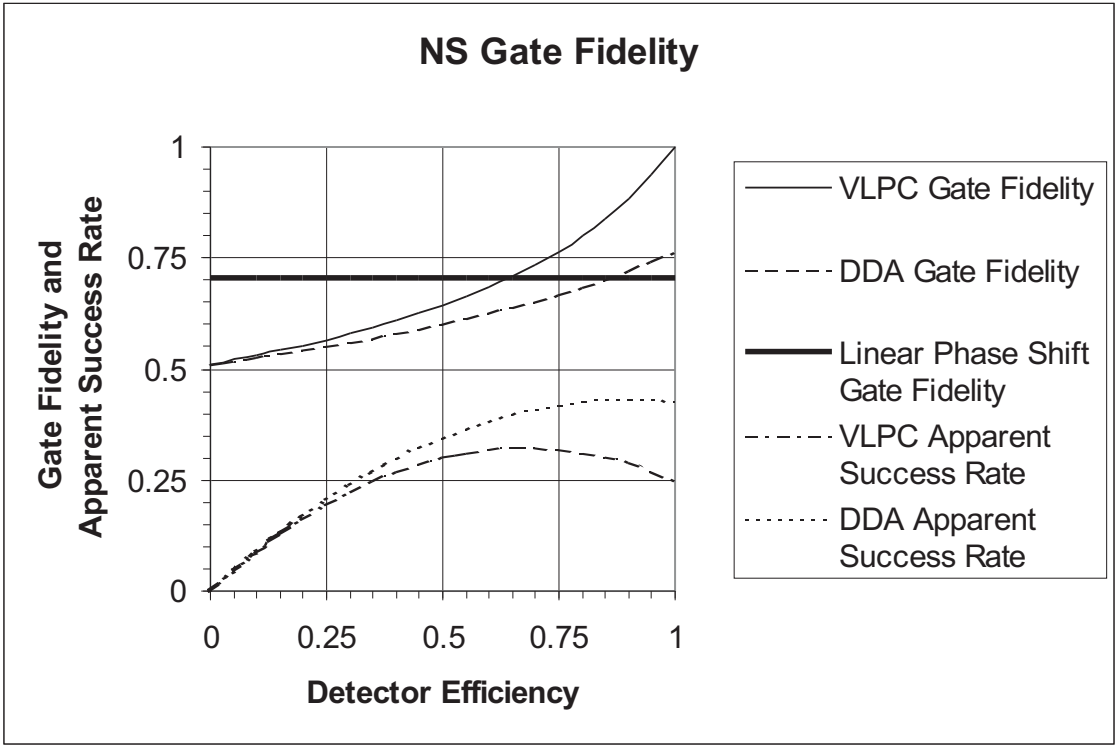}
  \caption{Gate fidelity as a function of detector
    efficiency for an NS gate employing DDAs and VLPCs.  For
    comparison, the gate fidelity of the linear phase shift is given.
    Also plotted are the apparent success rates for the two schemes.}
  \label{fig:Fidelity}
\end{figure}

Fig.~\ref{fig:Fidelity} gives a plot of gate fidelity as a function of
detector efficiency for both the DDA-- and the VLPC--based schemes.
Also plotted are the apparent success rates of the gates.  Note that,
using ideal photon counting, the apparent success rate should be
$1/4$.  This plot gives insight into why the gate fidelity is not
unity: the increase in the apparent success over $1/4$ implies that
the gate is reporting success when in fact it is performing
incorrectly.  Note that the DDA--based NS gate does not yield an
apparent success rate of $1/4$ even for unit efficiency; this
discrepency is due to the fact that a DDA cannot distinguish
one from two photons.

\subsection{Comparing the NS gate to a linear optics scheme}

It is necessary to establish a benchmark for the gate fidelity, in
order to determine the ability of the NS gate to perform a nonlinear
transformation.  One approach is to determine how such a NS gate
compares with a gate that uses only linear optics.  For example,
consider the tranformation
\begin{equation}
  \label{eq:LinearPhaseGate}
  \exp(i\tfrac{\pi}{2}\hat{a}^\dag \hat{a}): |\psi\rangle =
  \alpha|0\rangle + \beta|1\rangle + \gamma|2\rangle 
  \to |\psi_{\rm LP}\rangle 
  = \alpha |0\rangle + i \beta |1\rangle - \gamma |2\rangle \, .
\end{equation}
This transformation is a linear phase shift, implemented using a phase
shifter or a path--length delay.  Although it does not implement the
NS gate (due to the factor of $i$ on the $|1\rangle$ term), it does a
similar transformation and thus one can calculate the fidelity,
comparing this output state to the desired NS output $|\psi'\rangle$:
\begin{equation}
  \label{eq:FidelityLP}
  F(|\psi'\rangle,|\psi_{\rm LP}\rangle) = \sqrt{1 - 2|\beta|^2 + 2
  |\beta|^4} \, . 
\end{equation}
The gate fidelity for the linear phase shift is obtained (using the
minimizing state with $\beta= 1/\sqrt{2}$) to be $F_{\rm LP} =
1/\sqrt{2}$, and it can easily be shown that this gate fidelity is the
maximum obtainable using linear optics alone.  Thus, we can construct
a gate with gate fidelity $1/\sqrt{2}$ using only linear optics
(specifically, a linear phase shift).  This value sets a benchmark for
any implementation of a NS gate: the gate fidelity must exceed
$1/\sqrt{2}$ in order to be considered a nonlinear gate.

Observing the plot of Fig.~\ref{fig:Fidelity}, it is seen that the NS
gate fidelity using DDAs exceeds the threshold of $1/\sqrt{2}$ for
$\eta > 0.85$, whereas the VLPC--based scheme exceeds this threshold
for $\eta > 0.65$.

\section{Conclusions}
\label{sec:Conclusions}

As shown here, the function of a nondeterministic nonlinear gate (such
as the NS gate) employing photon counting depends critically on the
photon counting measurement model.  The two photon counting schemes
analyzed here, representing current technology, can only approximate
an ideal photon counting measurement.  As a result, the NS gate
employing realistic detection schemes can only approximate the desired
nonlinear transformation.

We have characterised the function of the NS gate with two realistic
photon counting models by the gate fidelity, which describes the
worst--case fidelity of an output state compared with the ideal output
state, normalized to an apparently successful operation.  With DDAs,
the NS gate achieves a maximum gate fidelity of $F_{\rm NS gate}
\simeq 0.77$ with unit efficiency detectors, and does not achieve the
critical value of $1/\sqrt{2}$ for realistic SPCM efficiencies.  Thus,
we conclude that a nonlinear transformation using a DDA--based NS gate
cannot be achieved (although we note that adding more beamsplitters
and SPCMs to form an array can increase the maximum gate fidelity).
By comparison, the VLPC achieves a gate fidelity of one for unit
efficiency, and performs well ($F_{\rm NS gate} > 1/\sqrt{2}$) for
realistic VLPC efficiencies of $\eta \simeq 90\%$~\cite{Tak99}.  Note,
however, that the gate fidelity drops rapidly with efficiency less
than one.

Performing nonlinear transformations using only linear optics and
photon counting represents an exciting new paradigm for nonlinear
quantum optics and quantum information.  Current photodetection
devices such as the VLPC can allow for nondeterministic nonlinear
transformations to be performed using present technology.  However,
new and innovative photon counting schemes must be developed in order
to meet the stringent requirements of fault--tolerant gates.

\acknowledgments     
 
This project has been supported by an Australian Research Council
Large Grant.  SDB acknowledges the support of a Macquarie University
Research Fellowship and a Macquarie University New Staff Grant.  ED
acknowledges the support of a Stanford Graduate Fellowship.


\end{document}